\documentclass[11pt,a4paper]{article}
\usepackage{jcappub}
\usepackage{euscript,epsfig,amsmath,amssymb}
\usepackage{amsfonts,latexsym}
\usepackage{enumerate}

\usepackage{hyperref}
\usepackage{url}

\usepackage[table]{xcolor}
\usepackage{tabularx}
\usepackage{multirow}
\usepackage[normalem]{ulem}
\usepackage{lineno}
\usepackage{afterpage}
\usepackage[labelsep=period,font=small]{caption}
\makeatletter

\definecolor{tableShade}{gray}{0.9}

\newcommand{\nn}{\nonumber}

\newcommand{\chic}{\chi_\textrm{c}} 

\newcommand*\patchAmsMathEnvironmentForLineno[1]{%
  \expandafter\let\csname old#1\expandafter\endcsname\csname #1\endcsname
  \expandafter\let\csname oldend#1\expandafter\endcsname\csname end#1\endcsname
  \renewenvironment{#1}%
     {\linenomath\csname old#1\endcsname}%
     {\csname oldend#1\endcsname\endlinenomath}}%
\newcommand*\patchBothAmsMathEnvironmentsForLineno[1]{%
  \patchAmsMathEnvironmentForLineno{#1}%
  \patchAmsMathEnvironmentForLineno{#1*}}%
\AtBeginDocument{%
\patchBothAmsMathEnvironmentsForLineno{equation}%
\patchBothAmsMathEnvironmentsForLineno{align}%
\patchBothAmsMathEnvironmentsForLineno{flalign}%
\patchBothAmsMathEnvironmentsForLineno{alignat}%
\patchBothAmsMathEnvironmentsForLineno{gather}%
\patchBothAmsMathEnvironmentsForLineno{multline}%
}

\newcommand{\vast}{\bBigg@{3}}
\newcommand{\Vast}{\bBigg@{5}}

\newcolumntype{L}[1]{>{\hsize=#1\hsize\raggedright\arraybackslash}X}%
\newcolumntype{R}[1]{>{\hsize=#1\hsize\raggedleft\arraybackslash}X}%
\newcolumntype{C}[1]{>{\hsize=#1\hsize\centering\arraybackslash}X}%

\renewcommand{\arraystretch}{1.2}

\makeatother

\begin{document}

\title{Cosmic infinity: A dynamical system approach}

\author[a,b,c,d]{Mariam~Bouhmadi-L\'opez,}
\author[a,b]{Jo\~{a}o~Marto,}
\author[c]{Jo\~ao~Morais}
\author[b,e]{and C\'{e}sar~M.~Silva}

\affiliation[a]{Departamento de F\'{i}sica, Universidade da Beira Interior\\
Rua Marqu\^{e}s D'\'{A}vila e Bolama, 6201-001 Covilh\~{a}, Portugal}
\affiliation[b]{Centro de Matem\'{a}tica e Aplica\c{c}\~{o}es da Universidade da Beira Interior (CMA-UBI)\\
Rua Marqu\^{e}s D'\'{A}vila e Bolama, 6201-001 Covilh\~{a}, Portugal}
\affiliation[c]{Department of Theoretical Physics, University of the Basque Country
UPV/EHU\\ P.O. Box 644, 48080 Bilbao, Spain}
\affiliation[d]{IKERBASQUE, Basque Foundation for Science\\ 48011, Bilbao, Spain}
\affiliation[e]{Departamento de Matem\'{a}tica, Universidade da Beira Interior\\
Rua Marqu\^{e}s D'\'{A}vila e Bolama, 6201-001 Covilh\~{a}, Portugal}

\emailAdd{mbl@ubi.pt}
\emailAdd{jmarto@ubi.pt}
\emailAdd{jviegas001@ikasle.ehu.eus}
\emailAdd{csilva@ubi.pt}

\abstract{
Dynamical system techniques are extremely useful to study cosmology. It turns out that in most of the cases, we deal with finite isolated fixed points corresponding to a given cosmological epoch. However, it is equally important to analyse the asymptotic behaviour of the universe. On this paper, we show how this can be carried out for 3-forms model. In fact, we show that there are fixed points at infinity mainly by introducing appropriate compactifications and defining a new time variable that washes away any potential divergence of the system. The richness of 3-form models allows us as well to identify normally hyperbolic non-isolated fixed points. We apply this analysis to three physically interesting situations: (i) a pre-inflationary era; (ii) an inflationary era; (iii) the late-time dark matter/dark energy epoch.
}


\arxivnumber{1611.03100}

\maketitle

\flushbottom


%
%
\section{Introduction}
\label{Introduction}

In recent years,  p-form cosmology \cite{Germani:2009iq,Koivisto:2009sd,Koivisto:2009fb,Koivisto:2009ew} has been the focus of a renewed interest as possible alternatives to scalar fields in explaining the nature of early inflation and the current dark energy (DE) fuelled acceleration. In particular, 3-forms, which in 4-dimensional space-times have one degree of freedom, have attracted a lot of attention since they can achieve a de Sitter phase without the need of slow-roll  conditions \cite{Duff1980}.

Dynamical systems have long since been a very useful tool in cosmology \cite{Wainwright2005,Coley2003}, in particular, in studying the evolution of cosmological models with scalar fields where explicit solutions of the evolution equations cannot usually be obtained.
In the case of 3-forms, since the early work by Koivisto and Nunes \cite{Koivisto:2009fb}, a dynamical system approach has been employed to study cosmological models with 3-forms. This includes works on 3-form inflation \cite{DeFelice:2012jt,Kumar2014} and re-heating \cite{DeFelice:2012wy}, models with interaction between cold dark matter (CDM) and a 3-form field playing the role of DE \cite{Ngampitipan:2011se,Boehmer:2011tp,2013PhRvD..88l3512K,Morais:2016bev}, or 3-form fields in a Randall-Sundrum II braneworld scenario \cite{Barros:2015evi}.

In systems where the dynamical  variables are unbounded, a compactification scheme \cite{Jordan2007a,Zhang2006,Gingold2004,Elias2006,Gingold2013}, which maps the system to a compact space, is required in order to understand the flow of the system in the vicinity of infinity. In refs.~\cite{Boehmer:2011tp,Morais:2016bev} a compact dynamical system description for models with CDM and 3-form DE was employed. This method allowed an intuitive identification of the fixed points of the system at infinite values of the 3-form field \cite{Morais:2016bev}. Among these, two repulsive solutions were found, which correspond to matter dominated epochs in the asymptotic past of the system.
On the present paper, we propose an adequate mathematical compactification that indeed proves that our intuitive results for those two points are correct \cite{Morais:2016bev}. This new compactification not only makes the 3-form compactified variable finite, as was already the case in \cite{Morais:2016bev}, but also allows us to obtain proper finite eigenvalues for the fixed points at infinite values of the 3-form field. The method used, developed in \cite{Elias2006}, consists in parametrising the independent variable in a certain direction preserving way and then considering a certain compactification chosen from a set of admissible compactifications. This procedure maps finite critical points to finite critical points and allows us to define critical point at infinity in a  way that is independent of the parametrisation. This method can be extremely important when willing to characterise the asymptotic behaviour of the Universe and one of the intrinsic degrees of freedom reaches extremely large values, making a dynamical system analysis subtle unless proper care is taken on the process.

This paper is structured as follows:
in section~\ref{3-Form cosmology} we review the cosmological model of a homogeneous and isotropic universe filled with CDM and a 3-form field playing the role of DE and its usual dynamical system description.
In section~\ref{New compact description} we review how to apply a compactification scheme to properly identify the fixed points at infinity and introduce a new dynamical system description of the model with different compact variables.
In section~\ref{Gaussian potential} we apply this new description to the particular case of a 3-form with a Gaussian potential and identify and characterise the fixed points at infinite values of the 3-form field.  We discuss the impact of initial de Sitter phases in the context of early inflation.
 In section~\ref{Pre-inflationary Universe: matter era} we analyse a 3-form model that leads to a pre-inflationary matter era followed by a de Sitter-like inflation.
In the concluding section~\ref{Discussion and Conclusions}, we discuss the results obtained and draw conclusions on the strengths and limitations of the method developed, as well as the possibility to extend it to a more general class of models.

%
%
\section{3-Form cosmology}
\label{3-Form cosmology}

\subsection{Cosmological model}

Let us consider the Friedmann-Lema\^itre-Robertson-Walker (FLRW) metric
\begin{align}
	\label{metric}
	ds^2 = -dt^2 + a^2(t)\delta_{ij}d x^i d x^j 
	\,,
\end{align}
which describes a spatially flat, homogeneous and isotropic Universe. In eq.~\eqref{metric}, $t$ is the cosmic time, $a(t)$ is the scale factor and $x^i$, $i=1,2,3$, are the comoving spatial coordinates.  In such geometry the only dynamical component of a massive 3-form $A_{\mu\nu\rho}$ \cite{Koivisto:2009sd} is the spatial-like component $A_{ijk}$, which can be parametrised  in terms of a scalar quantity $\chi(t)$ as \cite{Koivisto:2009sd,Koivisto:2009fb,Koivisto:2009ew}
\begin{align}
	\label{chi_definition}
	A_{ijk} = a^3(t)\chi(t)\epsilon_{ijk}
	\,.
\end{align} 
Here, the symbol $\epsilon_{ijk}$ is $+1$ $(-1)$ if $\{ijk\}$ is an even (odd) permutation of $\{123\}$ and $0$ if otherwise. In the action introduced in \cite{Koivisto:2009sd}, the potential of the 3-form is an arbitrary function $V(A_{\mu\nu\rho}A^{\mu\nu\rho})$. With the ansatz \eqref{chi_definition} we can write $A_{\mu\nu\rho}A^{\mu\nu\rho}=6\chi^2$, which means that the potential can be written as $V(\chi^2)$.
The evolution equation of $\chi(t)$ reads \cite{Koivisto:2009sd,Koivisto:2009fb,Koivisto:2009ew}
\begin{align}
	\label{chi_EquationOfMotion}
	\ddot\chi + 3H\dot\chi + 3\dot{H}\chi + 2\chi\frac{\partial V}{\partial\chi^2} = 0
	\,,
\end{align}
where $H:=\dot{a}/a$ is the Hubble parameter.

In this work, we consider a cosmological model for the late-time Universe with a 3-form field, playing the role of DE, and CDM. The Friedmman and Raychaudhuri equations read
\begin{align}
	\label{Friedmann}
	H^2 = \frac{\kappa^2}{3}\left[\rho_m + \frac{1}{2}\left(\dot\chi + 3H\chi\right)^2 + V\right]
	\,,
	\qquad
	\dot{H} =-\frac{\kappa^2}{2}\left(\rho_m + 2\chi^2 \frac{\partial V}{\partial\chi^2}\right)
	\,,
\end{align}
where $\kappa^{2}=8\pi G$, $G$ is the gravitational constant, and $\rho_m$ is the energy density of CDM, which satisfies the usual conservation equation $\dot\rho_m + 3H\rho_m=0$.

\begin{figure}
\centering
\includegraphics[width=0.666\columnwidth]{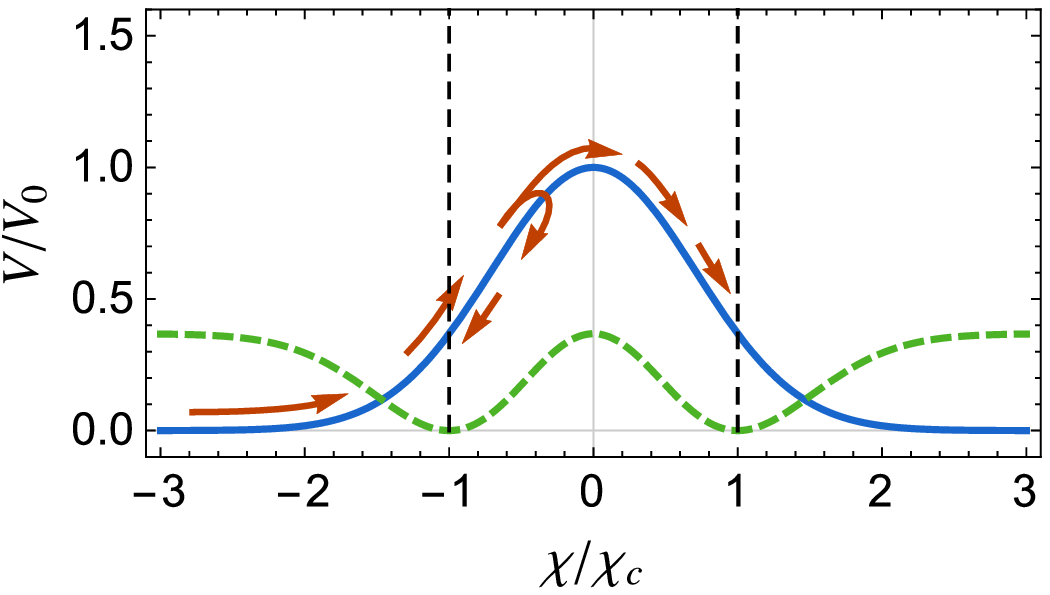}%
\caption{\label{fig:Gaussian_Potential} Representation of the evolution of the scalar quantity $\chi$. Independently of the shape of the potential, the field $\chi$ decays until it enters the interval $[-\chic,\,\chic]$. Once inside, the field $\chi$ evolves towards a minimum of the potential or the extremal points of the interval $[-\chic,\,\chic]$. In the case of the Gaussian potential $V=V_0 \exp[-\xi(\chi/\chic)^2]$ (blue line) represented here for $\xi=1$, with $V_0$ a positive constant, the field decays to one of the points $\chi=\pm\chic$ which acts as a local minimum of the effective potential (green curve), defined implicitly by $\partial V^\textrm{eff}/\partial\chi=(1-\chi^2/\chic^2)\partial V/\partial\chi$ \cite{Koivisto:2009fb}. For simplicity, we have fixed an arbitrary constant on the definition of $V^\textrm{eff}$ such that the effective potential vanishes at minima.}
\end{figure}

As discussed in Ref.~\cite{Morais:2016bev} a particularity of cosmological models with 3-forms is that, independently of the shape of the potential, the field $\chi$ decays monotonically until it enters the interval $[-\chic,\,\chic]$, with \mbox{$\chic :=\sqrt{2/(3\kappa^2)}$}. Once inside, it evolves towards a minimum of the potential; since the field cannot escape this interval (cf. figure~\ref{fig:Gaussian_Potential}), the limiting points $\chi=\pm\chic$ can act as local extrema. In particular, if $\partial V/\partial \chi^2(\pm\chic)<0$, the limiting points can act as local attractors that lead the system towards a Little Sibling of the Big Rip (LSBR) event \cite{Bouhmadi-Lopez:2014cca} in the asymptotic future \cite{Morais:2016bev}, i.e., an event at infinite cosmic time where $a$ and $H$ blow up while $\dot{H}$ remains finite.

\subsection{Dynamical system description}

A dynamical systems approach was first employed in the context of 3-form cosmology in Ref.~\cite{Koivisto:2009fb}.
There, three compact variables related to the fractional energy density of CDM and to the fractional kinetic and potential energy densities of the 3-form were defined
\footnote{Here, we prefer the letter $s$ for the variable related to the CDM energy density, in detriment of the letter $w$ used in \cite{Koivisto:2009fb}.}%
\begin{align}
	\label{System1_syz_definition}
	s :=\sqrt{\frac{\kappa^2 \rho_m}{3H^2}}
	\,,
	\quad
	y :=\frac{\dot\chi+3H\chi}{3H\chic }
	\,,
	\quad
	z :=\sqrt{\frac{\kappa^2 V}{3H^2}}
	\,,
\end{align}
with $0\leq s\leq1$, $-1\leq y\leq1$, and $0\leq z\leq1$. With these definitions, the Friedmann equation \eqref{Friedmann} can be re-written as 
\begin{align}
	\label{Friedmann_constraint}
	1 = s^2 + y^2 + z^2
	\,,
\end{align}
which allows the elimination of one dynamical variable from the system.
An autonomous dynamical system can be obtained by defining a fourth variable $\chi$
\begin{align}
	\label{System1_x_definition}
	x  := \frac{\chi}{\chic}
	\,,
\end{align}
where we employ a slightly different rescaling from the one used in Ref.~\cite{Koivisto:2009fb} so that the critical values $\chi=\pm\chic$ correspond to $x=\pm1$. Using the variables \eqref{System1_syz_definition} and \eqref{System1_x_definition} we can write the evolution equations for a cosmological model with CDM and a 3-form field as
\begin{align}
	\label{System1_Eqx}
	x' =&~ 3(y-x)
	\,,
	\\
	\label{System1_Eqy}
	y' =&~ \frac{1}{2}\left[
		3y\left(1-y^2-z^2\right)
		+\left(1- xy\right)\lambda(x)z^2
	\right]
	\,,
	\\
	\label{System1_Eqz}
	z' =&~ \frac{1}{2}z\left[
		3\left(1-y^2-z^2\right)
		-\left(y-x+x z^2\right)\lambda(x)
	\right]
	\,,
	\\
	s=&~\sqrt{1-y^2-z^2}
	\,.
\end{align}
Here, a prime indicates a derivative with respect to \mbox{$N:=\log(a/a_0)$} and the factor $\lambda(x)$ is defined as
\begin{align}
	\label{System1_lambda}
	\lambda(x) := -\sqrt{\frac{6}{\kappa^2}} \frac{1}{V}\frac{\partial V}{\partial \chi} = -3\frac{1}{V}\frac{\partial V}{\partial x}
	\,.
\end{align}

The set of eqs.~\eqref{System1_Eqx}-\eqref{System1_lambda} defines the evolution of the system in a subspace $\mathcal{M}$ of the three-dimensional space $(x,\,y,\,z)$ that corresponds to a half-cylinder  of radius unity and infinite height: $-1\leq y\leq1$, $0\leq z\leq\sqrt{1-y^2}$, and $-\infty<x<+\infty$. The borders of the half-cylinder
\begin{align}
	\label{B0_def}
	\mathcal{M}_0 =&~ \left\{ \left(x,\,y,\,z\right)\in\mathbb{R}^3 \, : \, z=0 \, \wedge \, -1\leq y \leq 1 \right\}
	\,, 
	\nn\\
	\mathcal{M}_1 =&~ \left\{ \left(x,\,y,\,z\right)\in\mathbb{R}^3 \, : \, y^2+z^2=1 \, \wedge \, z \geq 0 \right\}
	\,,
\end{align}
represent two invariant subsets of the system: the plane $\mathcal{M}_0$ corresponds to a model with CDM and a massless 3-form, which has been found to be equivalent to $\Lambda$CDM \cite{Koivisto:2009fb}; the surface $\mathcal{M}_1$ represents a universe filled solely by a 3-form field \cite{DeFelice:2012jt}.

The fixed points of the system~\eqref{System1_Eqx}-\eqref{System1_lambda} were classified in Ref.~\cite{Koivisto:2009fb} into three different categories: a saddle point A with $(x,\,y,\,z)=(0,0,0)$ that corresponds to an unstable matter era; two points B with $(x,\,y,\,z)=(\pm1,\pm1,0)$ that for some choices of the potential represent late-time attractors that lead the Universe to a LSBR event \cite{Morais:2016bev}; a set of points C corresponding to local extrema of the potential within the interval $[-\chic,\,\chic]$ and which can be either attractors or saddle points.
In refs.~\cite{Boehmer:2011tp,Morais:2016bev} the variable $x$ was replaced by the compact variable
\footnote{This compact variable was first proposed in Ref.~\cite{Boehmer:2011tp} where it was identified by the letter $x$. To avoid a potential confusion with the nomenclature, we adopt the letter $u$ for the compact variable, as used in Ref.~\cite{Morais:2016bev}}%
\begin{align}
	\label{System1_compactification}
	u := \frac{2}{\pi}\arctan\left(x\right)
	\,,
\end{align}
with $-1\leq u\leq1$. This substitution allowed for the identification of the fixed points at infinite values of the field $\chi$ \cite{Morais:2016bev}, i.e. for $u=\pm1$. Some of the new fixed points at infinite $\chi$ were found to correspond to the asymptotic past of the system. These points are characterised by their extremely repulsive nature \cite{Morais:2016bev}.

\section{New compact description}
\label{New compact description}

\subsection{Compactification}

When a compactification scheme is employed, like the one in \eqref{System1_compactification},  some terms may appear in the new evolution equations that diverge as the old variables approach infinity. When this happens, the dynamical system obtained after the compactification can be written as \cite{Jordan2007a,Zhang2006,Gingold2004,Elias2006,Gingold2013}
\begin{align}
	\begin{bmatrix}
	u'
	\\
	y'
	\\
	z'
	\end{bmatrix}
	=
	\frac{1}{g(u)}
	\begin{bmatrix}
	f_1\left(u,\,y,\,z\right)
	\\
	f_2\left(u,\,y,\,z\right)
	\\
	f_3\left(u,\,y,\,z\right)
	\end{bmatrix}
	\,,
\end{align}
where $g(u)$ vanishes as $u\rightarrow\pm1$. The divergence carried by $g(u)$ can then be washed away by defining a new time variable $\tau$, $d\tau = g^{-1}(u)\,d  N$ \cite{Gingold2004,Elias2006,Gingold2013}, such that the previous system can be written as
\begin{align}
	\frac{\partial }{\partial \tau} 
	\begin{bmatrix}
	u
	\\
	y
	\\
	z
	\end{bmatrix}
	=
	\begin{bmatrix}
	f_1\left(u,\,y,\,z\right)
	\\
	f_2\left(u,\,y,\,z\right)
	\\
	f_3\left(u,\,y,\,z\right)
	\end{bmatrix}
	\,.
\end{align}
The fixed points at \mbox{$\chi$-infinity} can now be identified as the points $(\pm1,\,y,\,z)$ such that $f_1=f_2=f_3=0$ \cite{Gingold2004,Elias2006,Gingold2013}.
A correct identification of these fixed points depends on whether or not the function $g(u)$ carries the divergent leading order of the equations, so that all the divergent terms are cancelled through a proper redefinition of the time variable. If this is not the case, we run the risk of ``overshooting'' in the divergence cancellation and introduce artificial fixed points in the system.

In \cite{Morais:2016bev} the proper identification and characterisation of the fixed points at \mbox{$\chi$-infinity} encountered three main difficulties. First, the fact that a trigonometric, instead of a polynomial, relation was employed in eq.~\eqref{System1_compactification} makes it more difficult to identify the divergence rate of the equations. Secondly, the fact that $z$ depends on $u$ through the potential, means that one has to take special care and understand what is the behaviour of $z$ as $u\rightarrow\pm1$. For potentials that vanish at infinite $\chi$ the variable $z$ may tend to $0$ sufficiently fast and cancel the divergent terms in $u$, e.g. the case of the Gaussian potential which was extensively discussed in Ref.~\cite{Morais:2016bev}. In such cases, a ``blind'' time redefinition based only on the divergent terms in $u$ would lead to incorrect results when identifying the fixed points of the system. Finally, even after the correct fixed points with $u=\pm1$ were identified, their stability needed to be clarified. Within the approach we will present next, we will see how to redefine the proper conditions to use a linear stability analysis.

\subsection{New dynamical system}
\label{New dynamical system}

We now present an alternative dynamical system description which tries to avoid the issues mentioned above.
We begin by employing a new compactification scheme for $\chi$
\begin{align}
	\label{v_def}
	\frac{\chi}{\chic} = \frac{v}{1-v^2}
	\,,
\end{align}
with $v\in[-1,\,1]$.
Note that the values $v=\pm1$ correspond to $\chi\rightarrow\pm\infty$. By employing the relation \eqref{v_def}, we ensure that all the divergent terms appear as powers of $(v\pm1)$ with negative exponents, facilitating the identification of the leading order of the divergence.

In a second step, we decompose the variable $z$ and isolate its explicit dependence on $v$. To do this, we first re-scale the potential $V$ as
\begin{align}
	\label{V_rescaling}
	V(v) = \frac{3H_0^2}{\kappa^2}V_*(v)
	\,,
\end{align}
where $H_0$ is the current value of the Hubble parameter and $V_*$ is a dimensionless function of $v$. Next, we introduce the compact Hubble rate
\footnote{%
A similar compact variable related to the Hubble rate is introduced in \cite{Alho:2015cza}.%
}
\begin{align}
	\label{h_def}
	h := \frac{\left(H/H_0\right)^2}{1+\left(H/H_0\right)^2}
	\,,
\end{align}
defined in the interval $[0,\,1]$, with $h=0$ corresponding to a Minkowski space-time, $H=0$, and $h=1$ to the limit
\footnote{We are assuming expanding cosmologies, i.e. $H\geq0$.}
$H\rightarrow+\infty$. Using eqs.~\eqref{V_rescaling} and \eqref{h_def} we can write the variable $z$ as
\begin{align}
	\label{z2_vyh}
	z^2 = \frac{1-h}{h}V_*(v)
	\,.
\end{align}

We are now in a position to write the set of evolution equations for the dynamical variables $(v,\,y,\,h)$. From eqs.~\eqref{Friedmann}, \eqref{System1_Eqx}, \eqref{System1_Eqy}, \eqref{v_def} and \eqref{z2_vyh}, we obtain
\begin{align}
	\label{System3_Eqv}
	v' =&~ 3\frac{1-v^2}{1+ v^2}\left[y\left(1-v^2\right)-v\right]
	\,,
	\\
	\label{System3_Eqy}
	y' =&~ \frac{3}{2}\Bigg\{
		y\left(1-y^2\right)
		-\frac{1-h}{h}
		\left[
			V_*(v) y
			+\frac{1-v^2}{1+ v^2}\frac{\partial V_*}{\partial v}\left(1-v^2- vy\right)
		\right]
	\Bigg\}
	\,,
	\\
	\label{System3_Eqh}
	h' =&~-3\left(1-h\right)\Bigg[
		h\left(1-y^2\right)
		+ \left(1-h\right)
		\left(
			\frac{1-v^2}{1+ v^2} v\frac{\partial V_*}{\partial v}
			-V_*(v)
		\right)
	\Bigg]
	\,.
\end{align}
For each type of potential, we can replace $V_*(v)$ in eqs.~\eqref{System3_Eqy} and \eqref{System3_Eqh}, identify the leading divergent term in order to proceed with the appropriate time redefinition and divergence cancellation, and finally identify the fixed points at infinity and study their stability. 
From eqs.~\eqref{Friedmann_constraint} and \eqref{z2_vyh} we can write $s$ as
\begin{align}
	\label{New_Friedmann_Constraint}
	s^2 = 1 - y ^2 -  \frac{1-h}{h}V_*(v)
	\,.
\end{align}

To conclude this section, we look at the invariant sets $\mathcal{M}_0$ and $\mathcal{M}_1$, cf. \eqref{B0_def}, in this new description. First, we note that the set $\mathcal{M}_0$ is no longer present in the system for a general potential. Instead, the behaviour of the system in $\mathcal{M}_0$ is given by the set of equations \eqref{System3_Eqv}, \eqref{System3_Eqy} and \eqref{System3_Eqh} with the null potential $V_*=0$.
In the case of $\mathcal{M}_1$, i.e. in the absence of CDM, the combination of eq.~\eqref{z2_vyh} with the condition $z^2+y^2=1$ allows us to express $h$ as
\begin{align}
	\label{noCDM_h}
	h = \frac{V_*(v)}{1-y^2+V_*(v)}
	\,.
\end{align}
As such, we can drop eq.~\eqref{System3_Eqh} and re-write eqs.~\eqref{System3_Eqv} and \eqref{System3_Eqy} as
\begin{align}
	\label{System3_Eqv_noCDM}
	v' =~ 3\frac{1-v^2}{1+ v^2}\left[y\left(1-v^2\right)-v\right]
	\,,
	\quad
	y' =~ -\frac{3}{2}\left(1-y^2\right)
		\frac{1-v^2}{1+ v^2}\frac{1}{V_*}\frac{\partial V_*}{\partial v}\left(1-v^2- vy\right)
	\,.
\end{align}
These equations are equivalent to the ones obtained in Ref.~\cite{DeFelice:2012jt} in the context of 3-form inflation.


\section{Gaussian potential}
\label{Gaussian potential}

For concreteness, from now on we analyse the case of the Gaussian potential
\begin{align}
	\label{gaussian_def}
	V = \frac{3H_0^2}{\kappa^2}\bar{V} \,e^{-\xi \frac{\chi^2}{6\kappa^2}}
	\,,
\end{align}
which was studied in detail in Ref.~\cite{Morais:2016bev}. Here, $\xi$ and $\bar{V}$ are positive dimensionless constants.  With the ansatz \eqref{gaussian_def} we can write $V_*$ and its derivative in terms of $v$ as
\begin{align}
	\label{Gaussian_V*}
	V_* = \bar{V} \exp\left[-\frac{\xi}{9}\left(\frac{v}{1-v^2}\right)^2\right]
	\,,
	\qquad
	\frac{\partial V_*}{\partial v} = -\frac{2\xi}{9}\bar{V} \frac{\left(1+v^2\right)v}{\left(1-v^2\right)^3} \exp\left[-\frac{\xi}{9}\left(\frac{v}{1-v^2}\right)^2\right]
	\,.
\end{align}
In order to completely identify the fixed points of the system at $\chi$-infinity, we will first analyse the solutions in the subset $\mathcal{M}_1$, when only a 3-form field is present, and then in the more general case, where a 3-form field and CDM are present.

\subsection{FLRW with a 3-form: Early Universe}
\label{Gaussian: Only 3-form}

We begin by analysing the case of a Universe filled solely by a 3-form field with a Gaussian potential. Such models, where a 3-form is considered without other kinds of matter, have been considered before, in particular in the context of 3-form inflation \cite{Koivisto:2009fb,DeFelice:2012jt,Kumar2014} with the 3-form playing the role of the inflaton.
In this case, the set of equations that govern the evolution of the system is \eqref{System3_Eqv_noCDM}. After replacing $V_*$ and its derivative in these equations, we find that the leading order of the divergent terms that appear in the equation for $y'$ in eq.~\eqref{System3_Eqv_noCDM} is $(1-v^2)^{-2}$. We absorb these divergences by defining a new time variable $\tau$, with \mbox{$d\tau =(1+v^2)^{-1}(1-v^2)^{-2} dN$},  such that the new dynamical system reads
\footnote{%
Strictly speaking, only the factor $(1-v^2)^{2}$ is necessary in order to wash away the divergence of the system. The factor $(1+v^2)$ is included in order to simplify the final equations.
}%
\begin{align}
	\label{System4_Eqv_noCDM}
	\frac{\partial v}{\partial \tau} = 3\left(1-v^2\right)^{3}\left[y\left(1-v^2\right)-v\right]
	\,,
	\qquad
	\frac{\partial y}{\partial \tau} = \frac{\xi}{3}\left(1-y^2\right)
		\left(1-v^2- vy\right) \left(1+v^2\right)v
	\,.
\end{align}
This system has nine fixed points: $\hat{\mathcal{B}}^\pm$, $\hat{\mathcal{C}}$, $\hat{\mathcal{D}}_{+1}^\pm$, $\hat{\mathcal{D}}_{0}^\pm$ and $\hat{\mathcal{D}}_{-1}^\pm$. The coordinates $(v_{fp},\,y_{fp})$ and other characteristics of the fixed points are presented in table~\ref{FixedPoints_noCDM}, while in figure~\ref{fig:noCDM_stream} we show the flow and the position of fixed points.

\afterpage{
\noindent
\begin{minipage}{\textwidth}
\centering

\rowcolors{2}{}{gray!10}
\renewcommand{\arraystretch}{1.5}
\begin{tabularx}{\textwidth}{| C{1} C{1.4} C{0.5} C{1} C{1} C{0.55} C{.55} |}
\hiderowcolors 
\hline 
	\multirow{2}{*}{\centering Fixed point}
	&
	\multirow{2}{*}{$\left(v_{fp},\,y_{fp}\right)$}
	&
	\multirow{2}{*}{\centering$h_{fp}$}
	&
	\multirow{2}{\linewidth}{\centering Physical state}
	&
	\multirow{2}{*}{\centering Stability}
	&
	\multicolumn{2}{c|}{Type}
\\
	
	&
	
	&
	
	&
	
	&
	
	&
	\cite{Koivisto:2009fb}
	&
	\cite{Morais:2016bev}
\\
\hline\hline
	$\hat{\mathcal{B}}^\pm$
	&
	$\left(\pm\frac{\sqrt{5}-1}{2},\,\pm1\right)$
	&
	1
	&
	LSBR
	&
	Attractor
	&
	$B_\pm$
	&
	I
\\\hline
	$\hat{\mathcal{C}}$
	&
	$\left(0,\,0\right)$
	&
	$\frac{\bar{V}}{1+\bar{V}}$
	&
	dS
	&
	Saddle
	&
	$C$
	&
	II
\\\hline
	$\hat{\mathcal{D}}_{+1}^\pm$
	&
	$\left(\pm1,\,+1\right)$
	&
	undet.
	&
	dS
	&
	Repulsive
	&
	n/a
	&
	III
\\\hline
	$\hat{\mathcal{D}}_{0}^\pm$
	&
	$\left(\pm1,\,0\right)$
	&
	0
	&
	M
	&
	Saddle
	&
	n/a
	&
	III
\\\hline
	$\hat{\mathcal{D}}_{-1}^\pm$
	&
	$\left(\pm1,\,-1\right)$
	&
	undet.
	&
	dS
	&
	Repulsive
	&
	n/a
	&
	III
\\
\hline
\end{tabularx}\\
\captionof{table}{\label{FixedPoints_noCDM}%
Fixed points of the system \eqref{System4_Eqv_noCDM} for a cosmological model with a 3-form with a Gaussian potential \eqref{gaussian_def}. For each solution, we present the value of $h$ at the fixed point, the asymptotic physical state of the Universe: LSBR - Little Sibling of the Big Rip; dS - de Sitter; M - Minkowski; the stability and its classification according to refs.~\cite{Koivisto:2009fb} and \cite{Morais:2016bev}.
}

\vspace{15pt}

\centering
\includegraphics[width=0.5\columnwidth]{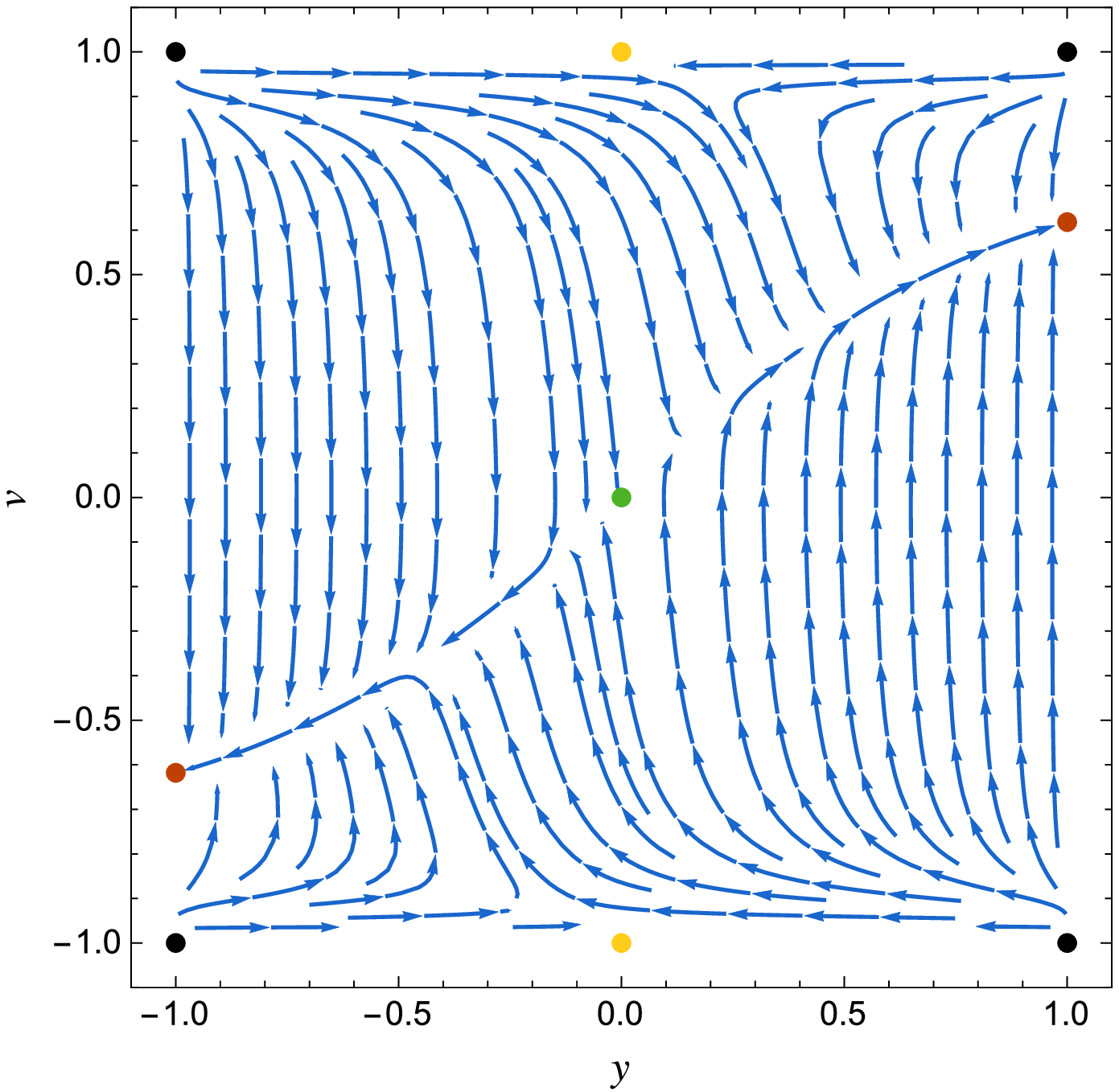}
\captionof{figure}{\label{fig:noCDM_stream}%
Flow (blue arrows) and fixed points (coloured dots) of the system \eqref{System4_Eqv_noCDM} for a cosmological model with a 3-form with a Gaussian potential \eqref{gaussian_def} with  $\xi=1$. The coordinates of the fixed points can be checked in table~\ref{FixedPoints_noCDM}. At \mbox{$\chi$-infinity} ($v=\pm1)$ the system has six fixed points: four repulsive points and two saddles.}
\vspace{20pt}

\end{minipage}
}

Of the nine fixed points obtained, $\hat{\mathcal{B}}^\pm$ and $\hat{\mathcal{C}}$ were already identified in \cite{DeFelice:2012jt} while the other six correspond to solutions at \mbox{$\chi$-infinity}: $\hat{\mathcal{D}}_{+1}^\pm$, $\hat{\mathcal{D}}_{0}^\pm$ and $\hat{\mathcal{D}}_{-1}^\pm$.
The eigenvalues, $\gamma_i$, of the respective Jacobians are
\begin{align}
	\label{eigenvalues_noCDM_1}
	\left\{\gamma_1,\,\gamma_2\right\}_{\hat{\mathcal{D}}_{+1}^\pm} = \left\{0,\,\frac{4}{3}\xi\right\}
	\,,
	\quad
	\left\{\gamma_1,\,\gamma_2\right\}_{\hat{\mathcal{D}}_{0}^\pm} = \left\{0,\,-\frac{2}{3}\xi\right\}
	\,,
	\quad
	\left\{\gamma_1,\,\gamma_2\right\}_{\hat{\mathcal{D}}_{-1}^\pm} = \left\{0,\,\frac{4}{3}\xi\right\}
	\,.
\end{align}
In all six cases we encounter a null eigenvalue. Therefore, to study the behaviour of the system in the vicinity of each fixed point,  we cannot use the results of linear stability analysis and we must instead resort to the methods of Centre Manifold Theory (CMT) \cite{Carr1981,Rendall2002,Boehmer:2011tp}.

We begin by applying the double transformation
\begin{align}
	\label{noCDM_double_transformation}
	\begin{bmatrix}
		v\\
		y
	\end{bmatrix}
	\,\rightarrow \,
	\begin{bmatrix}
		\delta v\\
		\delta y
	\end{bmatrix}
	=
	\begin{bmatrix}
		 v-v_{fp}\\
		 y- y_{fp}
	\end{bmatrix}
	\,\rightarrow \,
	\begin{bmatrix}
		 X_1\\
		 X_2
	\end{bmatrix}
	=M^{-1}
	\begin{bmatrix}
		\delta v\\
		\delta y
	\end{bmatrix}
	\,,
\end{align}
where $M$ is the square matrix whose columns are the eigenvectors of the Jacobian of the system at $(v_{fp},\,y_{fp})$. In the new coordinate system $(X_1,\,X_2)$, the fixed point sits at the origin $(0,\,0)$, the axes are aligned with the eigenvectors of the Jacobian, and the set of equations \eqref{System3_Eqv_noCDM} reads
\begin{align}
	\label{System5_noCDM}
	\frac{\partial }{\partial\tau}
	\begin{bmatrix}
		X_1\\
		X_2
	\end{bmatrix}
	= 
	\begin{bmatrix}
		0&0\\
		0&\lambda_2
	\end{bmatrix}
	\cdot
	\begin{bmatrix}
		X_1\\
		X_2
	\end{bmatrix}
	+
	\begin{bmatrix}
		F_1\left(X_1,\,X_2\right)\\
		F_2\left(X_1,\,X_2\right)
	\end{bmatrix}
\end{align}
where $F_{1,2}\sim O(||(X_1,\,X_2)||^2)$.
At this point, we note that CMT \cite{Carr1981} guarantees the existence of a centre manifold $W^c$, tangent to $\vec{e}_1=(1,\,0)$ at the origin.%
\footnote{In addition to the centre manifold $W^c$, tangent to $\vec{e}_2=(0,\,1)$ at the origin, the system has a stable manifold $W_s$ whenever $\Re(\lambda_2)<0$ and an unstable manifold $W_u$ whenever $\Re(\lambda_2)>0$. In the case of a Gaussian potential \eqref{gaussian_def} with positive $\xi$ the fixed points $\hat{\mathcal{D}}_{+1}^\pm$ and $\hat{\mathcal{D}}_{-1}^\pm$ are repulsive in the direction $\vec{e}_2$ while the fixed points $\hat{\mathcal{D}}_{0}^\pm$ are attractive, cf. Eqs~\eqref{eigenvalues_noCDM_1}.
}
In addition, it assures that in some neighbourhood of the origin there is a mapping $G(X_1)$, such that \mbox{$(X_1,\, G(X_1))\in W^c$}. This mapping must satisfy \mbox{$G(0)=0$} and \mbox{$(\partial G/\partial \,X_1)(0)=0$}, and can be Taylor expanded around $X_1=0$ up to an arbitrary order as
\begin{align}
	\label{G expansion}
	G\left(X_1\right) = A_2\, X_1^2 + A_3\, X_1^3 + \dots
\end{align}
The main result of CMT is that, in a vicinity of the origin, the stability of the system in $W^c$ can be decided through the stability of the reduced system
\begin{align}
	\label{noCDM_reduced}
	\frac{\partial X_1}{\partial \tau} = F_1\left(X_1,\,G\left(X_1\right)\right)
	\,.
\end{align}

We are now left with the task of finding the values of the linear coefficients $A_n$. For this, we the combine the second row of eq.~\eqref{System5_noCDM} with eq.~\eqref{noCDM_reduced} and the Leibnitz rule, \mbox{$\partial X_2/\partial \tau =(\partial G/\partial X_1) (\partial X_1/\partial \tau)$} in order to write
\begin{align}
	\frac{\partial G}{\partial X_1}F_1\left(X_1,\,G\left(X_1\right)\right) 
	= \lambda_2G\left(X_1\right) 
	+ F_2\left(X_1,\,G\left(X_1\right)\right)
		\,.
\end{align}
This equation can be expanded up to an arbitrary order in $X_1$ by means of eq.~\eqref{G expansion} and, after collecting terms of equal order in $X_1$, we can iteratively compute  each of the linear coefficients $A_n$.

When we apply this method to the fixed points $\hat{\mathcal{D}}_{+1}^\pm$, $\hat{\mathcal{D}}_{0}^\pm$ and $\hat{\mathcal{D}}_{-1}^\pm$, we find, for all the six cases,
\begin{align}
	\label{Gaussian_noCDM_finalv}
	\frac{\partial X_1}{\partial \tau} = 24 X_1^3 + O\left(X_1^4\right)
	\,.
\end{align}
Consequently, around all those fixed points the system is repulsive in the centre manifold.
Since $\xi>0$, we conclude from eq.~\eqref{eigenvalues_noCDM_1} that the two pairs $\hat{\mathcal{D}}_{+1}^\pm$ and $\hat{\mathcal{D}}_{-1}^\pm$ are repulsive and represent possible states in the asymptotic past of the system, while the pair $\hat{\mathcal{D}}_{0}^\pm$ consists of two saddle points. In addition, from eq.~\eqref{noCDM_h} we find that while for $\hat{\mathcal{D}}_{0}^\pm$ the Hubble parameter vanishes at the fixed point, for the fixed points $\hat{\mathcal{D}}_{+1}^\pm$ and $\hat{\mathcal{D}}_{-1}^\pm$ the value of $h$ is undetermined. This means that in the past, each trajectory begins in a de Sitter state as the 3-form field behaves like a cosmological constant of an arbitrary value. 

The fact that the same eq.~\eqref{Gaussian_noCDM_finalv} is obtained in all cases is not a coincidence. In fact, when we solve eq.~\eqref{Gaussian_noCDM_finalv}  to leading order and re-write the solution in terms of $\chi$,  we conclude that for very large values of $|\chi|$ the field $\chi$ behaves as $\chi\sim a^{-3}$. This result is in concordance with the behaviour obtained for $|\chi|\gg\chic$ from the analysis of the Friedmann equation, cf. Ref.~\cite{Morais:2016bev}, and with the tracking/scaling behaviour discussed in Ref.~\cite{Koivisto:2009ew}.

\subsection{On the possibility of an early de Sitter era}
\label{Application to inflation}

In the previous section we have seen that, in the absence of CDM, a 3-form with the Gaussian potential \eqref{gaussian_def} can lead to a de Sitter phase in the asymptotic past, with an arbitrary energy scale. However, despite the de Sitter behaviour and the flatness of the potential this initial phase is not characterised by a slow-roll regime -- in fact the field $\chi$ decays rapidly with $\chi\sim a^{-3}$.
Using the method developed in section~\ref{New dynamical system} we can generalise this result for other 3-form models in which the derivative of the potential vanishes asymptotically for high values of $|\chi|$, e.g. potentials that asymptotically behave as $V\sim V_{\infty}(1 - e^{-\xi \chi^2})$, $V\sim V_{\infty}e^{-\xi |\chi|}$, $V\sim V_{\infty}(1+ \xi|\chi|^{-n})$, \dots

This general result suggests the possibility of using these types of models to have the Universe starting in a de Sitter phase, thus avoiding the question of how the proper initial conditions for inflation are achieved. However, despite this appealing initial behaviour of the Gaussian potential (and others), the model might be characterised by some instabilities due to the negativeness of the squared speed of sound of the 3-form, $c_s^2 = \chi V_{\chi\chi}/V_{\chi}$ \cite{Koivisto:2009fb,Morais:2016bev}. In fact, for example in the case of the Gaussian potential, even though there is always an interval of values of $\chi$ around the origin for which the squared speed of sound is positive and the parameters of the model can be tuned in order for this interval is arbitrarily large so that the field $\chi$ rapidly passes its border, as we come sufficiently close to the repulsive fixed point, i.e. as we approach the asymptotic past, there is always a point at which the squared speed of sound becomes negative.
A way to avoid this issue is to assume a phase transition that occurs at a finite value of $\chi_\textrm{trans}$, such that in the past, for $|\chi|>\chi_\textrm{trans}$, the potential vanishes (or becomes constant), and that only starting from some given epoch, for $|\chi|<\chi_\textrm{trans}$, the Gaussian potential enters into play.

\begin{figure}[t]
\centering
\includegraphics[width=0.317\textwidth]{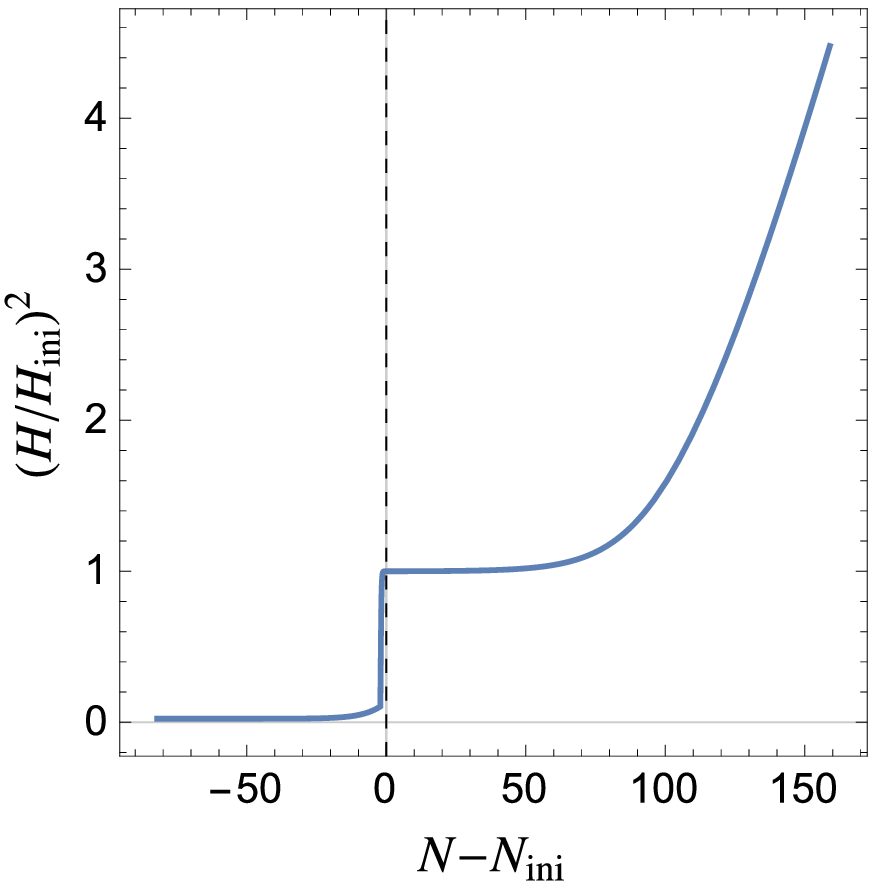}%
\hfill
\includegraphics[width=0.33\textwidth]{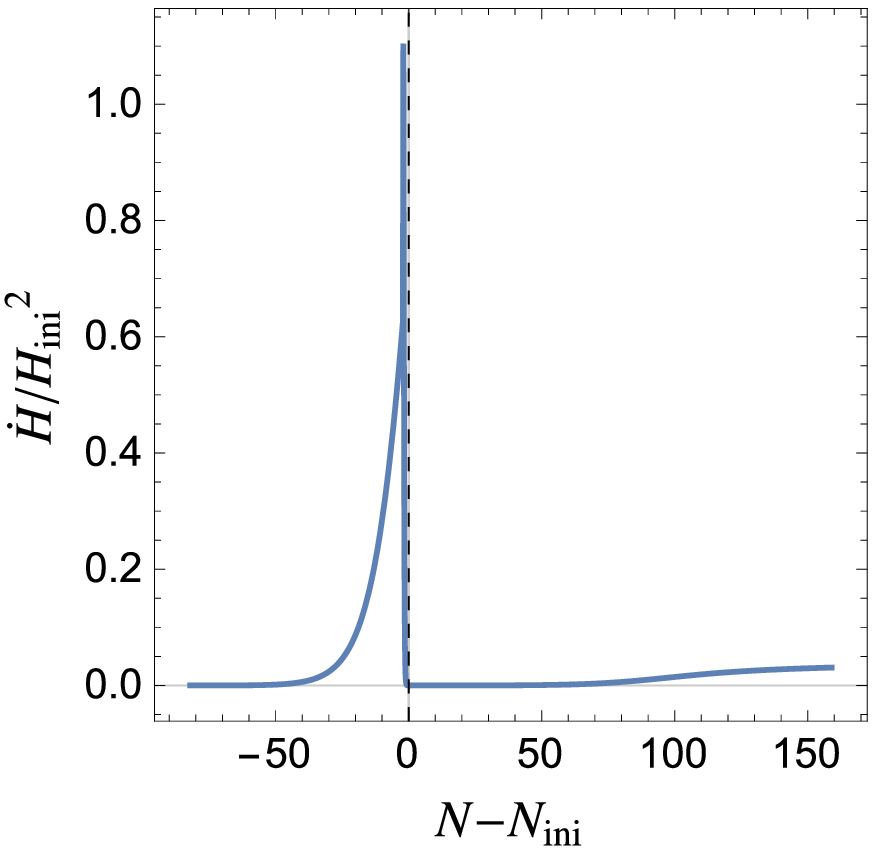}%
\hfill
\includegraphics[width=0.33\textwidth]{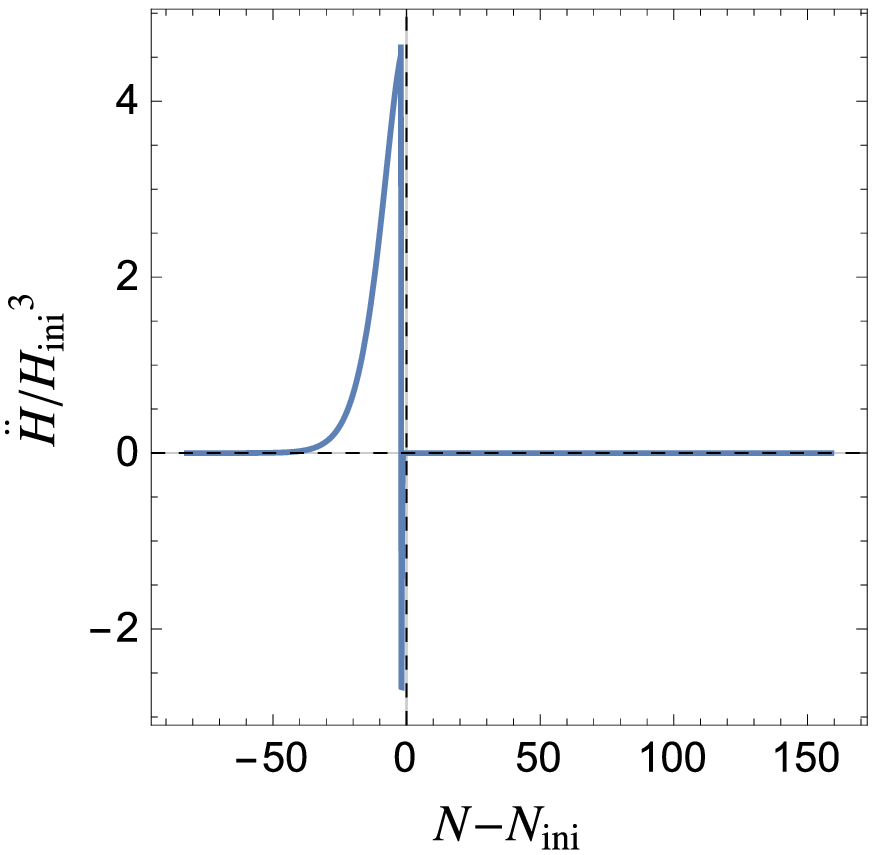}%
\caption{\label{figure_H_H2_H3}
The evolution of $H^2$, $\dot{H}$, and $\ddot{H}$ in the case of the Gaussian potential as functions of the number of e-folds, $N$. The reference value $N_{ini}$ is chosen to correspond to the moment the 3-form is at the top of the potential.
}
\end{figure}

In addition, we note that, although this initial de Sitter phase might prove unsuccessful in accounting for proper inflation, it is always followed by a second inflationary era which is reached as the field $\chi$ approaches the neighbourhood of the maximum of the potential and enters the slow-roll regime characterised by $\ddot\chi\simeq0$. This second de Sitter-like inflation can then be used to realise the necessary e-folds of inflation that set the seeds for the large scale structure of the Universe. The background evolution of such an epoch can be seen in figure~\ref{figure_H_H2_H3}, where $H^2$, $\dot{H}$ and $\ddot{H}$ are plotted as functions of the number of e-folds. After the initial de Sitter phase for $N-N_{ini}<0$ and a brief super inflationary period, a new de Sitter epoch sets in and begins around $N_{ini}$ and lasts for at least $\sim60$ e-folds. Since in the present case, i.e. for the Gaussian potential, we have $\dot{H}>0$ at all times, we find that this second inflationary period does not give place to a radiation epoch. Instead $\dot{H}$ tends to a positive non-vanishing constant and the Universe evolves towards a LSBR. Nevertheless, an effective mechanism of re-heating, usual and necessary in most inflationary models, can then be employed to give place to a radiation era once the necessary e-folds of inflation are obtained.


\subsection{FLRW with a 3-form and CDM: Late Universe}
\label{subsubsec: With CDM}

\afterpage{
\noindent
\begin{minipage}[t]{\textwidth}
\centering

\rowcolors{2}{}{gray!10}
\centering
\renewcommand{\arraystretch}{1.5}
\begin{tabularx}{\columnwidth}{|C{1.2} C{1.6} C{.5} C{1.1} C{1.4} C{.6} C{.6}|}
\hiderowcolors\hline
	
	\multirow{2}{\linewidth}{\centering Fixed point}
	&
	\multirow{2}{*}{$\left(v_{fp},\,y_{fp},\,h_{fp}\right)$}
	&
	\multirow{2}{*}{$s_{fp}$}
	&
	\multirow{2}{\linewidth}{\centering Physical state}
	&
	\multirow{2}{\linewidth}{\centering Stability}
	&
	\multicolumn{2}{c|}{Type}
\\
	
	&
	
	&
	
	&
	
	&
	
	&
	\cite{Koivisto:2009fb}
	&
	\cite{Morais:2016bev}
\\
\hline\hline
	${\mathcal{A}}$
	&
	$\left(0,\,0,\,1\right)$
	&
	1
	&
	CDM
	&
	Saddle
	&
	A
	&
	I
\\\hline
	${\mathcal{B}}^\pm$
	&
	$\left(\pm\frac{\sqrt{5}-1}{2},\,\pm1,\,1\right)$
	&
	0
	&
	LSBR
	&
	Attractor
	&
	$B_\pm$
	&
	I
\\\hline
	${\mathcal{C}}$
	&
	$\left(0,\,0,\,\frac{\bar{V}}{1+\bar{V}}\right)$
	&
	0
	&
	dS
	&
	Saddle
	&
	$C$
	&
	II
\\\hline
	${\mathcal{E}_0^\pm}$
	&
	$\left(\pm1,\,0,\,1\right)$
	&
	1
	&
	CDM
	&
	Repulsive
	&
	n/a
	&
	III
\\\hline
	${\mathcal{F}}_{+1}^\pm$
	&
	$\left(\pm1,\,+1,\,h\right)$
	&
	0
	&
	dS
	&
	Saddle
	&
	n/a
	&
	III
\\\hline
	${\mathcal{F}}_{-1}^\pm$
	&
	$\left(\pm1,\,-1,\,h\right)$
	&
	0
	&
	dS
	&
	Saddle
	&
	n/a
	&
	III
\\
\hline
\end{tabularx}
\captionof{table}{\label{FixedPoints_CDM}%
Fixed points of the system \eqref{System6_Eqv}-\eqref{System6_Eqh} for a cosmological model with a 3-form with a Gaussian potential \eqref{gaussian_def} and CDM. For each solution we present the value of $s$ at the fixed point, the asymptotic physical state of the Universe: CDM - cold dark matter domination; LSBR - Little Sibling of the Big rip; dS - de Sitter; M - Minkowski; the stability and the classification according to refs.~\cite{Koivisto:2009fb} and \cite{Morais:2016bev}.
}

\vspace{15pt}

\includegraphics[width=.49\columnwidth]{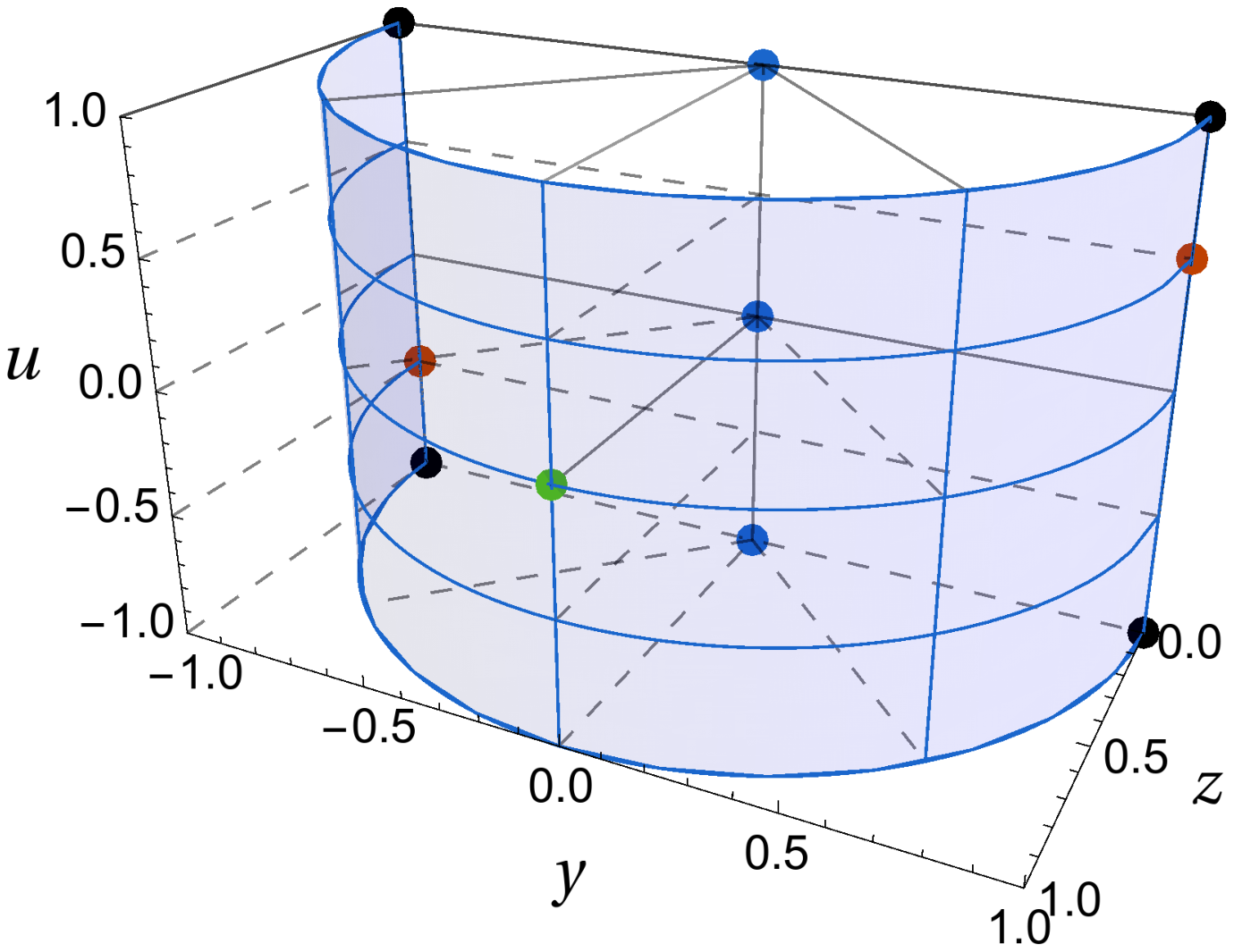}
\hfill
\includegraphics[width=.49\columnwidth]{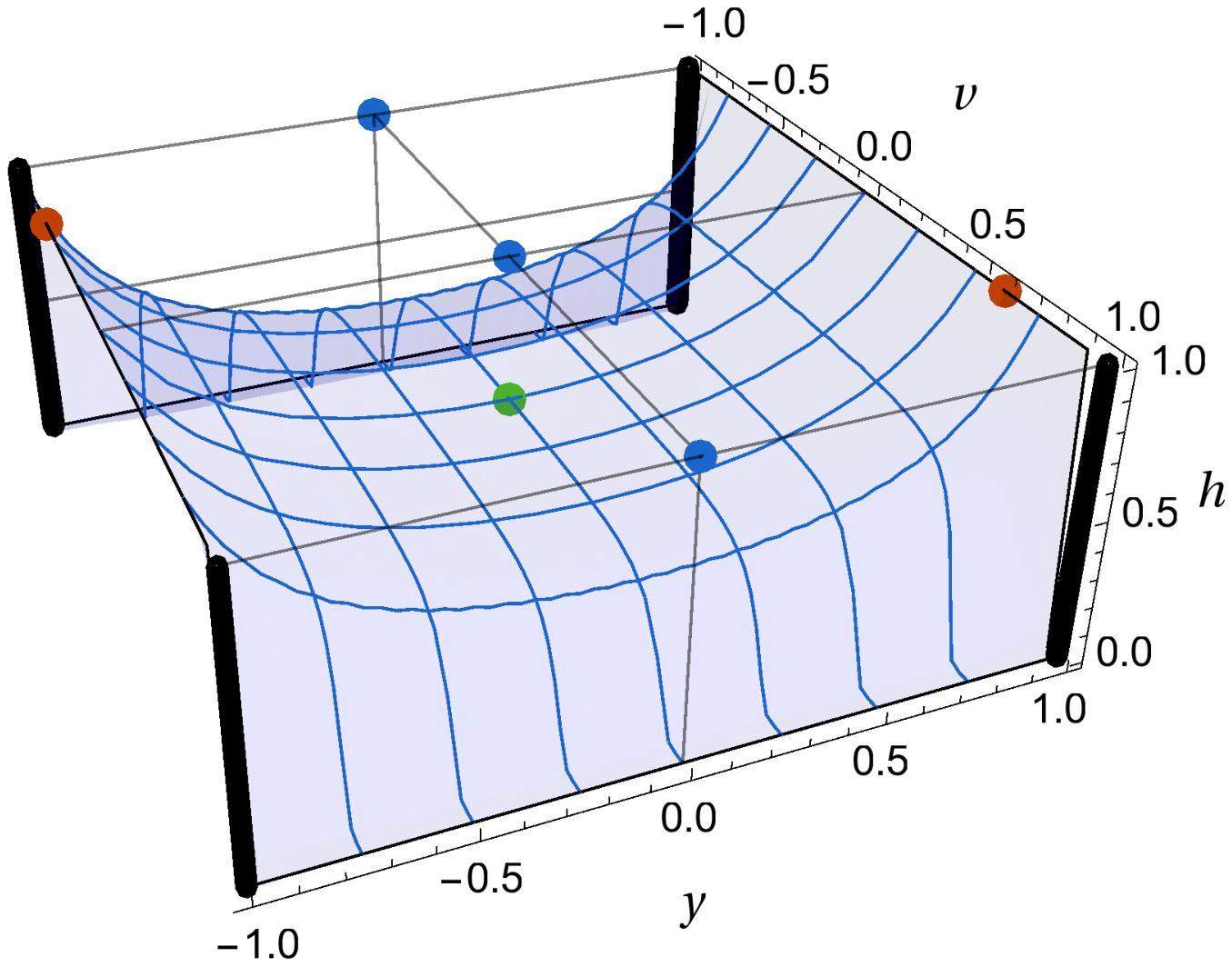}
\captionof{figure}{\label{fig:CDM_fixedpoints}%
Position of the fixed points of the cosmological model with a 3-form field with a Gaussian potential (see eq.~\eqref{gaussian_def} with $\xi=\bar{V}=1$) and CDM in: (left panel) the dynamical system description $(u,\,y,\,z)$ of refs.~\cite{Boehmer:2011tp,Morais:2016bev}; (right panel) the dynamical system description $(v,\,y,\,h)$ employed in this work. Corresponding fixed points are signalled with the same colour in both panels. In each panel the subset $\mathcal{M}_1$ corresponding to the case of no CDM is identified by the blue surface. The Friedmann constraint imposes that the physical system evolves within the half-cylinder on the left panel and above the surface $\mathcal{M}_1$ on the right panel. The coordinates $(v_{fp},\,y_{fp},\,h_{fp})$ of each point are presented in table~\ref{FixedPoints_CDM}.
}
\vspace{20pt}

\end{minipage}
}

We now focus our attention on cosmological models with CDM and a 3-form field with a Gaussian potential, i.e. late-time cosmological models with a 3-form playing the role of DE \cite{Koivisto:2009fb,Ngampitipan:2011se,Boehmer:2011tp,2013PhRvD..88l3512K,Morais:2016bev}.
By substituting \eqref{Gaussian_V*} in eqs.~\eqref{System3_Eqv}, \eqref{System3_Eqy} and \eqref{System3_Eqh} we obtain
\begin{align}
	\label{System6_Eqv}
	&~v' = 3\frac{1-v^2}{1+ v^2}\left[y\left(1-v^2\right)-v\right]
	\,,
	\\
	\label{System6_Eqy}
	&~y' = \frac{3}{2}\vast\{
		y\left(1-y^2\right)
		-\bar{V}\frac{1-h}{h}\exp\left[-\frac{\xi}{9}\left(\frac{v}{1-v^2}\right)^2\right]
		\left[
			  y 
			- \frac{2\xi}{9} \frac{v\left(1-v^2- vy\right)}{\left(1-v^2\right)^2} 
		\right]
	\vast\}
	\,,
	\\
	\label{System6_Eqh}
	&~h' =-3\left(1-h\right)\vast\{
		h\left(1-y^2\right)
		+ \bar{V} \left(1-h\right)
		\exp\left[-\frac{\xi}{9}\left(\frac{v}{1-v^2}\right)^2\right]
		\left[
			- \frac{2\xi}{9} \frac{v^2}{\left(1-v^2\right)^2}
			- 1
		\right]
	\vast\}
	\,,
\end{align}
with the constraint, derived from eq.~\eqref{New_Friedmann_Constraint},
\begin{align}
	0 \leq \left(1 - y^2\right)h -  \left(1-h\right)V_*(v) \leq 1
	\,.
\end{align}
Due to the fast decay of the potential, all terms in eqs.~\eqref{System3_Eqy} and \eqref{System3_Eqh} which depend on the quadratic exponential vanish for $v\rightarrow\pm1$ and no $v$-divergences exist in the system. In addition, we note that in the presence of CDM and for the Gaussian potential \eqref{gaussian_def} with positive $\xi$, the Hubble parameter never vanishes, i.e., $0<h\leq1$.

By inspection of eqs.~\eqref{System3_Eqv}, \eqref{System3_Eqy} and \eqref{System3_Eqh}, we find that the system has  six isolated fixed points: ${\mathcal{A}}$, ${\mathcal{B}}^\pm$, ${\mathcal{C}}$ and ${\mathcal{E}_0^\pm}$; and four sets of non-isolated fixed points: ${\mathcal{F}}_{+1}^\pm$ and ${\mathcal{F}}_{-1}^\pm$ (cf. table~\ref{FixedPoints_CDM} and the right panel of figure~\ref{fig:CDM_fixedpoints}). We will focus our attention on the stability of the system near the fixed points ${\mathcal{E}_0^\pm}$, ${\mathcal{F}}_{+1}^\pm$ and ${\mathcal{F}}_{-1}^\pm$, since these are the ones that correspond to states of the system at \mbox{$\chi$-infinity}.

For the pair $\mathcal{E}^\pm$, the Jacobian has three positive eigenvalues, $\gamma_i$,
\begin{align}
	\left\{\gamma_1,\,\gamma_2,\,\gamma_3\right\}_{\mathcal{E}^\pm}
	=
	\left\{\frac{3}{2},\,3,\,3\right\}
	\,.
\end{align}
indicating that the points are repulsive and represent possible states in the asymptotic past of the system. From eq.~\eqref{New_Friedmann_Constraint}, we obtain $s_{fp}^2=1$, therefore these points represent a past matter era.

Alternatively, for all the fixed points belonging to the sets $\mathcal{F}_{+1}^+=(1,\,1,\,h)$, $\mathcal{F}_{+1}^-=(1,\,-1,\,h)$, $\mathcal{F}_{-1}^+=(-1,\,1,\,h)$ and $\mathcal{F}_{-1}^-=(-1,\,-1,\,h)$, with $0<h\leq1$, the eigenvalues of the Jacobian are
\begin{align}
	\label{P1_eigenvalues}
	\left\{\gamma_1,\,\gamma_2,\,\gamma_3\right\}_{\mathcal{F}_{\pm1}^\pm}
	=
	\left\{3,\,-3,\,0\right\}
	\,.
\end{align}
The null eigenvalue corresponds to a direction tangent to the line containing the fixed points. Since the other two eigenvalues have non-zero real part, every point in the sets $\mathcal{F}_{\pm1}^\pm$ is said to be normally hyperbolic \cite{Aulbach1984,Coley2003} and the stability of the trajectories along the remaining directions can be determined by applying the usual linear stability theory. The existence of a positive and a negative eigenvalues lead us to conclude that the sets $\mathcal{F}_{\pm1}^\pm$ are composed entirely by saddle points which are neither attractive nor repulsive. A comparison with the results in section~\ref{Gaussian: Only 3-form} shows that the sets $\mathcal{F}_{\pm1}^\pm$ correspond to the fixed points $\hat{\mathcal{D}}_{\pm1}^\pm$ of table~\ref{FixedPoints_noCDM}.

%
%
\section{Pre-inflationary Universe: initial matter era}
\label{Pre-inflationary Universe: matter era}

\afterpage{
\noindent
\begin{minipage}{\textwidth}
\centering

\rowcolors{2}{}{gray!10}
\renewcommand{\arraystretch}{1.5}
\begin{tabularx}{\textwidth}{| C{1} C{1.4} C{0.5} C{1} C{1} C{0.55} C{.55} |}
\hiderowcolors 
\hline 
	\multirow{2}{*}{\centering Fixed point}
	&
	\multirow{2}{*}{$\left(v_{fp},\,y_{fp}\right)$}
	&
	\multirow{2}{*}{\centering$h_{fp}$}
	&
	\multirow{2}{\linewidth}{\centering Physical state}
	&
	\multirow{2}{*}{\centering Stability}
	&
	\multicolumn{2}{c|}{Type}
\\
	
	&
	
	&
	
	&
	
	&
	
	&
	\cite{Koivisto:2009fb}
	&
	\cite{Morais:2016bev}
\\
\hline\hline
	$\hat{\mathcal{B}}^\pm$
	&
	$\left(\pm\frac{\sqrt{5}-1}{2},\,\pm1\right)$
	&
	1
	&
	LSBR
	&
	Saddle
	&
	$B_\pm$
	&
	I
\\\hline
	$\hat{\mathcal{C}}$
	&
	$\left(0,\,0\right)$
	&
	$\frac{\bar{V}}{1+\bar{V}}$
	&
	dS
	&
	Attractor
	&
	$C$
	&
	II
\\\hline
	$\hat{\mathcal{D}}_{+1}^\pm$
	&
	$\left(\pm1,\,+1\right)$
	&
	1
	&
	undet.
	&
	Saddle
	&
	n/a
	&
	III
\\\hline
	$\hat{\mathcal{D}}_{0}^\pm$
	&
	$\left(\pm1,\,0\right)$
	&
	1
	&
	D
	&
	Repulsive
	&
	n/a
	&
	III
\\\hline
	$\hat{\mathcal{D}}_{-1}^\pm$
	&
	$\left(\pm1,\,-1\right)$
	&
	1
	&
	undet.
	&
	Saddle
	&
	n/a
	&
	III
\\
\hline
\end{tabularx}\\
\captionof{table}{\label{quasilinear_FixedPoints_noCDM}%
Fixed points of the system \eqref{quasilinear_Eqv_noCDM} for a cosmological model with a 3-form with a potential of the kind \eqref{quasi-linear_pot}. For each solution, we present the value of $h$ at the fixed point, the asymptotic physical state of the Universe: LSBR - Little Sibling of the Big Rip; dS - de Sitter; D - dust; the stability and its classification according to refs.~\cite{Koivisto:2009fb} and \cite{Morais:2016bev}.
}

\vspace{15pt}

\centering
\hspace{5pt}
\includegraphics[width=0.45\columnwidth]{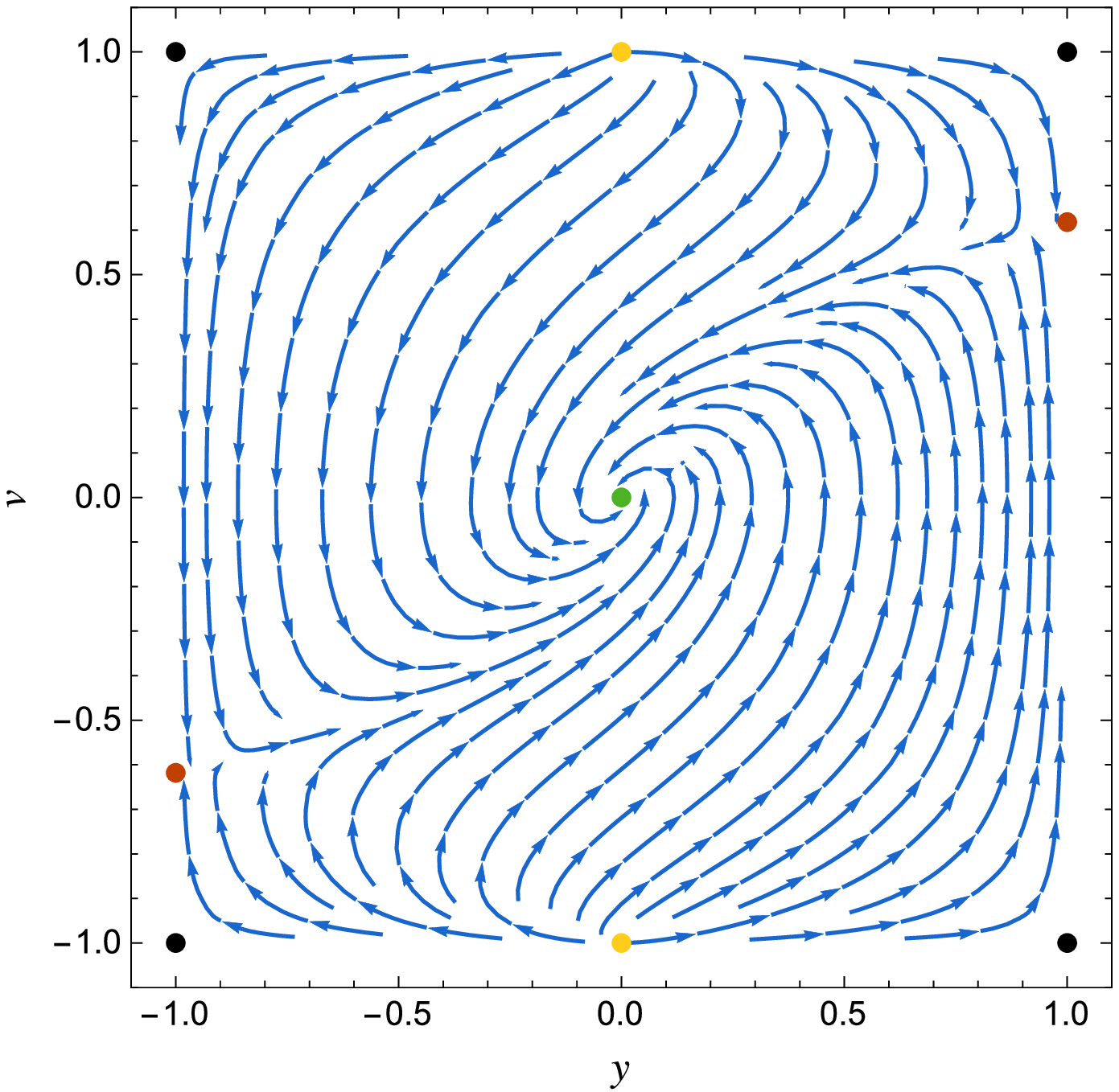}
\hfill
\includegraphics[width=0.45\columnwidth]{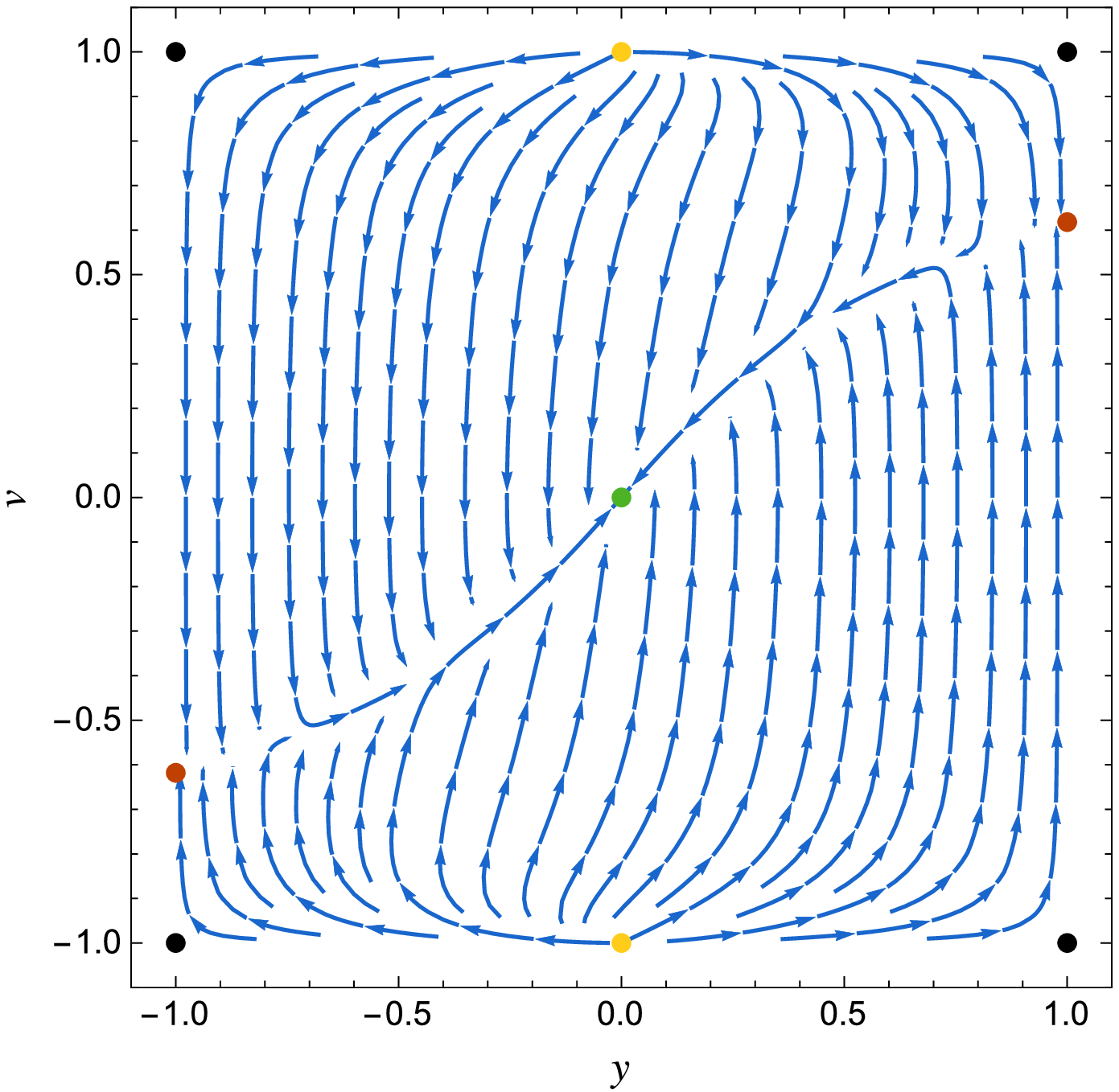}
\hspace{5pt}
\captionof{figure}{\label{fig:quasi_linearnoCDM_stream}%
Flow (blue arrows) and fixed points (coloured dots) of the system \eqref{quasilinear_Eqv_noCDM} for a cosmological model with a 3-form with a potential of the kind \eqref{quasi-linear_pot} with  $\xi=5$ (left panel) and $\xi =1/5$ (right panel). The coordinates of the fixed points can be checked in table~\ref{quasilinear_FixedPoints_noCDM}. At \mbox{$\chi$-infinity} ($v=\pm1)$ the system has six fixed points: four saddle points and repulsive solutions. For $\xi > 1/2$ the system starts to present a spiralling behaviour around the attractor $(0,0)$.}
\vspace{20pt}

\end{minipage}
}

In the previous section, we have analysed the case of a 3-form with a Gaussian potential which could be interesting in the context of early inflation and to describe the late-time universe dominated by CDM and DE, the latter being modelled by the 3-form. However, a 3-form scenario could also be suitable to describe a pre-inflationary era. In refs.~\cite{Powell:2006yg,Wang:2007ws,Scardigli:2010gm,BouhmadiLopez:2012by} it is discussed how an such era may leave pre-inflationary imprints on the primordial power spectrum which can alleviate the quadrupole anomaly in the CMB.
As an example of how a 3-form model could be interesting in this context we analyse the case of the potential
\begin{align}
	\label{quasi-linear_pot}
	V = V_0 \sqrt{1 + \xi\left(\frac{\chi}{\chic}\right)^2}
	\,,
\end{align}
where $V_0$ and $\xi$ are positive constants. For high values of the amplitude of the 3-form $(\chi\gg \chic/\sqrt\xi)$ the potential grows linearly with $\chi$. Therefore, if the potential term dominates in the past we expect the 3-form to behave effectively as dust in an initial era.%
\footnote{Here we remind the reader that for $|\chi|\gg\chic$ we expect the field $\chi$ to decay with $a^{-3}$ \cite{Koivisto:2009ew,Morais:2016bev}. Thus, in the region $|\chi|\gg\chic,\chic/\sqrt{\xi}$ the potential term behaves as a pressureless fluid.}%
As the value of $\chi$ decreases, the potential becomes increasingly flat, with a minimum $V_0$ at $\chi=0$.

As in the case of the Gaussian potential, we begin by writing the rescaled potential $V_*$ and its derivative in terms of $v$:
\begin{align}
	\label{quasi-linear}
	V_* 
	= \bar{V} \frac{\sqrt{\left(1-v^2\right)^2 + \xi v^2}}{1-v^2}
	\,,
	\qquad
	\frac{\partial V_*}{\partial v} 
	= 
	\bar{V}\xi \frac{v}{\sqrt{\left(1-v^2\right)^2 + \xi v^2}} \frac{1+v^2}{\left(1-v^2\right)^2}
	\,.
\end{align}
After substitution of eqs.~\eqref{quasi-linear} in eq.~\eqref{System3_Eqv_noCDM} we find the set of equations
\begin{align}
	\label{quasilinear_Eqv_noCDM}
	v' = 3 \frac{1-v^2}{1+v^2}\left[\left(1-v^2\right)y - v\right] 
	\,,
	\qquad
	y' = -\frac{3}{2}\xi \left(1-y^2\right)\frac{v\left(1-v^2 - vy\right)}{\left(1-v^2\right)^2 + \xi v^2}
	\,,
\end{align}
which does not present any divergences as $v\rightarrow\pm1$. This system has nine fixed points: $\hat{\mathcal{B}}^\pm$, $\hat{\mathcal{C}}$, $\hat{\mathcal{D}}_{+1}^\pm$, $\hat{\mathcal{D}}_{0}^\pm$ and $\hat{\mathcal{D}}_{-1}^\pm$. As in the case of the Gaussian potential, the points  $\hat{\mathcal{B}}^\pm$ and $\hat{\mathcal{C}}$ correspond to possible solutions in the asymptotic future of the system with a finite value of $\chi$, while the points $\hat{\mathcal{D}}_{+1}^\pm$, $\hat{\mathcal{D}}_{0}^\pm$ and $\hat{\mathcal{D}}_{-1}^\pm$ all have $|v|=1$ and therefore represent solutions at $\chi$-infinity that, when physically relevant, represent the asymptotic past. In table~\ref{quasilinear_FixedPoints_noCDM}, we present the coordinates of the fixed points and some of the  characteristics of the associated solutions, while in figure~\ref{fig:quasi_linearnoCDM_stream} we show the flow of the system and the position of the fixed points. Next, we present the stability analysis for each of these solutions.

Of the six fixed points that correspond to solutions at infinite values of $\chi$: $\hat{\mathcal{D}}_{+1}^\pm$, $\hat{\mathcal{D}}_{0}^\pm$ and $\hat{\mathcal{D}}_{-1}^\pm$; we find that the eigenvalues are:
\begin{align}
	\label{eigenvalues_quasilinear_infinite}
	\left\{\gamma_1,\,\gamma_2\right\}_{\hat{\mathcal{D}}_{+1}^\pm} = \left\{-3,\,3\right\}
	\,,
	\quad
	\left\{\gamma_1,\,\gamma_2\right\}_{\hat{\mathcal{D}}_{0}^\pm} = \left\{\frac{3}{2},\,3\right\}
	\,,
	\quad
	\left\{\gamma_1,\,\gamma_2\right\}_{\hat{\mathcal{D}}_{-1}^\pm} = \left\{-3,\,3\right\}
	\,.
\end{align}
We thus find that, contrary to what happened for the Gaussian potential, all the fixed points at $\chi-$infinity are hyperbolic. 
From the sign of the eigenvalues in \eqref{eigenvalues_quasilinear_infinite}, we conclude that the four solutions, ${\hat{\mathcal{D}}_{\pm1}^\pm}$, are unphysical saddle points while the points ${\hat{\mathcal{D}}_{0}^\pm}$ are repulsive and represent the past of the system. These repulsive solutions are characterised by a total domination of the potential term of the 3-form ($H^2\simeq \kappa^2V/3$) which confirms that with such a potential the 3-form induces a primordial dust dominated era.
In addition, for all the solutions at $\chi$-infinity we find the positive eigenvalue $3$ whose respective eigenvector is pointed in the direction of $\vec{e}_v$. This result shows once more that for $|\chi|\gg\chic$ the 3-form field behaves as $|\chi|\sim a^{-3}$.

The three other fixed points of the system: $\hat{\mathcal{B}}^\pm$ and $\hat{\mathcal{C}}$, represent solutions for finite values of the field $\chi$ ($v\neq\pm1$). Physically, these points can be interpreted as follows: while $\hat{\mathcal{B}}^\pm$, if attractive, will lead the Universe to a LSBR event in the future, the point $\hat{\mathcal{C}}$ corresponds to the case where the field decays to the minimum of the potential at $\chi=0$. The eigenvalues of such points are
\begin{align}
	\label{eigenvalues_quasilinear_finite}
	\left\{\gamma_1,\,\gamma_2\right\}_{\hat{\mathcal{B}}^\pm} = \left\{0,\,-3\right\}
	\,,
	\quad
	\left\{\gamma_1,\,\gamma_2\right\}_{\hat{\mathcal{C}}} 
	= \left\{
		-\frac{3}{2}\left(1-\sqrt{1-2\xi}\right)
		,\,
		-\frac{3}{2}\left(1+\sqrt{1-2\xi}\right)
	\right\}
	\,.
\end{align}
We thus find that the points ${\hat{\mathcal{B}}^\pm}$ have one null eigenvalue, which means that their stability cannot be decided through the linear stability analysis, while the point $\hat{\mathcal{C}}$ is hyperbolic for all values $\xi>0$. We next discuss the stability of such points.

We begin by the points ${\hat{\mathcal{B}}^\pm}$ which, due to the presence of the null eigenvalue require an analysis based on CMT. After applying the techniques discussed in section~\ref{Gaussian: Only 3-form}, we find that to leading order in $\delta y\equiv y - y_{fp}$
\begin{align}
	\delta y' \simeq \mp3\frac{\left(1-\sqrt{5}\right)^2}{3-\sqrt{5}}\frac{\xi}{1+\xi} \delta y^2
	\,,
\end{align}
where the negative sign is related to the behaviour near ${\hat{\mathcal{B}}^+}$ and the positive sign to the behaviour near ${\hat{\mathcal{B}}^-}$. Since $\delta y$ is negative near ${\hat{\mathcal{B}}^+}$ and positive near ${\hat{\mathcal{B}}^-}$ we find that in the centre manifold the system tends to increase the amplitude of small deviations $\delta y$,
meaning that the points ${\hat{\mathcal{B}}^\pm}$ are unstable saddles. Therefore, the potential \eqref{quasi-linear_pot} does not lead the system to a LSBR event in the future. This result is in concordance with the discussion in \cite{Morais:2016bev}; i.e. the occurrence of a LSBR depends on the sign of the derivative $\partial V/\partial \chi^2$ at the points $\chi=\pm\chic$ and only when $\partial V/\partial \chi^2(\chi=\pm\chic)<0$ does the LSBR occur. Indeed, in this case we have  $ \partial V/\partial \chi^2(\chi=\pm\chic)>0$ (cf. eq.~\eqref{quasi-linear_pot}).

On the other hand, the point ${\hat{\mathcal{C}}}$ is an hyperbolic de Sitter attractor with 2 eigenvalues with negative real part. Being ${\hat{\mathcal{C}}}$ the only attractor of the system, we conclude that in this model the Universe evolves towards a de Sitter inflationary era with $H^2=\kappa^2V_0/3$. This universal attractor can be reached in two distinct ways, depending on the value of $\xi$: for $\xi>1/2$ the eigenvalues acquire an imaginary part and the trajectories of the system enter a spiralling motion as they converge to the attractor ${\hat{\mathcal{C}}}$; for $\xi<1/2$ the eigenvalues are purely real and this oscillatory motion disappears. This difference in behaviour can be understood as follows: for $\xi<1/2$ the potential is extremely flat near the minimum and therefore the field $\chi$ approaches the end solution following the slow-roll condition $\ddot\chi\simeq0$ and no oscillations take place. For $\xi>1/2$, however, the potential is sufficiently concave for the field to reach the minimum with enough kinetic energy to overshoot it and give rise to the damped oscillations around the final state. These two different behaviours can be seen in figure~\ref{fig:quasi_linearnoCDM_stream}.

Summarising, in this section we have analysed the global dynamics of a 3-form model with the potential \eqref{quasi-linear_pot} and found that the 3-form behaves initially as a pressureless fluid and later, as the field rolls down to the minimum of the potential, ends in a de Sitter inflationary phase. This behaviour can be seen in figure~\ref{fig:quasi_linearnoCDM_stream} as all the trajectories start in the past matter points ${\hat{\mathcal{D}}_{0}^\pm}$ (yellow) and end in the universal de Sitter attractor at the origin  ${\hat{\mathcal{C}}}$ (green). As such, this model can drive, in a natural way, a pre-inflationary matter era that ends in a de Sitter inflation. Such era could be helpful in alleviating the quadrupole problem of the CMB, as discussed in \cite{Powell:2006yg,Wang:2007ws,Scardigli:2010gm,BouhmadiLopez:2012by}. Although outside the scope of this work, we pretend to carry out the analysis of the cosmological perturbations of such a model, and their implications in the CMB, in the near future. As a final note, we point out that unlike the models presented in section~\ref{Gaussian: Only 3-form}, for the potential \eqref{quasi-linear_pot} the squared speed of sound of the 3-form is always positive and smaller than or equal to unity, avoiding therefore any potential instability, including the absence of superluminal issues.

%
%
\section{Discussion and Conclusions}
\label{Discussion and Conclusions}

In our previous paper \cite{Morais:2016bev}, we employed a dynamical system approach to study late-time cosmologies with CDM and DE, the latter modelled by a massive 3-form field. and we identified six new fixed points which were previously disregarded in the literature. These new  solutions correspond to states of the system in the limit of infinite values of the 3-form field.
In the present work, we further explore the mathematical tools necessary for analysing the flow of a system in the vicinity of infinity. Through the use of a new dynamical system description and compactification scheme \cite{Jordan2007a,Zhang2006,Gingold2004,Elias2006,Gingold2013}, we were able to unequivocally identify the fixed points of the system in a model with a 3-form with a Gaussian potential, both in the presence and absence of non-interacting CDM. Subsequently, using linear stability theory, Centre Manifold Theory \cite{Carr1981,Rendall2002,Boehmer:2011tp}, and the notion of normally hyperbolic fixed points \cite{Aulbach1984,Coley2003}, we deduced the stability of the fixed points obtained.

In absence of CDM, we found six fixed points at \mbox{$\chi$-infinity} (cf. table~\ref{FixedPoints_noCDM}), of which two are saddles and four are repulsive. The latter correspond to the past asymptotic state of the system and represent an initial de Sitter inflationary epoch. The value of the Hubble parameter during this inflationary era is not determined by the system, instead it is characteristic of each individual trajectory. We discuss the implication of this initial de Sitter solution in the context of primordial inflation.
Alternatively, in the presence of CDM we obtained a pair of isolated fixed points, $\mathcal{E}^\pm$, and four connected sets, ${\mathcal{F}_{\pm1}^\pm}$, of non-isolated fixed points corresponding to states at \mbox{$\chi$-infinity} (cf. table~\ref{FixedPoints_CDM}). A detailed stability analysis shows that the two isolated solutions are repulsive in nature and correspond to matter dominated eras in the asymptotic past, while the four sets ${\mathcal{F}_{\pm1}^\pm}$ are composed of normally hyperbolic saddle points.  These results corroborate the findings of Ref.~\cite{Morais:2016bev}, where two repulsive, $\pi_0^\pm$, and four saddle, $\pi_{\pm1}^\pm$, fixed points were found at \mbox{$\chi$-infinity}. The difference in dimensionality of the manifolds of equilibria $\pi_0^\pm$ and ${\mathcal{F}_{\pm1}^\pm}$ can be explained by noting that the dynamical system employed in Ref.~\cite{Morais:2016bev} does not track the value of the Hubble parameter and therefore does not distinguish between points with different values of $h$. We note that the method presented in this paper allows us to compute the eigenvalues of the fixed points at $\chi$-infinity while avoiding the difficulties encountered in Ref.~\cite{Morais:2016bev}.

Although the results presented here focus on the case of a 3-form with a Gaussian potential, the set of variables $(v,\,y,\,h)$ introduced in this work can be easily applied to other types of potentials. 
As an example, we include the case of the quasi-linear potential $V=V_0\sqrt{1+\xi(\chi/\chic)^2}$ and show how, in the absence of other matter fields, the 3-form leads the Universe from an initial pre-inflationary era corresponding to a dust-like epoch to a final de Sitter solution. This behaviour could prove interesting in the context of modelling a pre-inflationary era, which could alleviate the quadrupole problem in the CMB power-spectrum \cite{Powell:2006yg,Wang:2007ws,Scardigli:2010gm,BouhmadiLopez:2012by}, a possibility we will explore in the future.
Exceptions for the applicability of this dynamical system representation are the potentials with zeros at finite values of $\chi$, as the dynamical system \eqref{System3_Eqv}-\eqref{System3_Eqh}, is not appropriate to study fixed-points with $h=0$. This limitation is reminiscent of the fact that the original system $(x,\,y,\,z)$ is not well defined for such potentials \cite{Koivisto:2009fb}. 
In addition, the fact that a separate compactification strategy is employed for each degree of freedom may generate an unclear interpretation when studying trajectories where more than one degree of freedom diverges. A way to overcome both these inconvenient situations is to go back to the original degrees of freedom of the model $(H,\,\rho_m,\,\chi,\, \dot\chi)$ and apply a compactification scheme of the kind discussed in \cite{Gingold2004,Elias2006,Gingold2013}. In principle, this would allow for a  complete dynamical system description of cosmological models with 3-forms that could be applied to a general potential. Finally, we note that such an approach could be extended to models with a scalar field which, contrary to a 3-form, can have fixed points at infinite values of the scalar field even in the asymptotic future. This will be explored in a future work \cite{Bouhmadi-Lopez:2017aaa}.

\acknowledgments

The Authors are grateful to Juan~M.~Aguirregabiria for enlightening discussions on dynamical system analysis.
The work of MBL is supported by the Portuguese Agency “Funda\c{c}\~ao para a Ci\^encia e Tecnologia” through an Investigador FCT Research contract, with reference IF/01442/2013/ CP1196/CT0001. MBL and JMorais wish to acknowledge the support from the Basque government Grant No. IT592-13 (Spain), FONDOS FEDER under grant FIS2014-57956-P (Spanish government) and the COST Action CA15117 (CANTATA).
JMorais is also thankful to UPV/EHU for a PhD fellowship.
This research work is supported by the grant UID/MAT/00212/2013.



\bibliographystyle{JHEP}
\bibliography{References}

\end{document}